\documentclass[preprint,prb,preprintnumbers,amsmath,amssymb]{revtex4}

\usepackage{txfonts}
\usepackage{graphicx}
\usepackage{bm}
\usepackage{color}
\usepackage{times} %Times New Roman on

\begin{document}
\title{Non-uniform magnetic system driven by non-magnetic ion substitution \\ in CaRu$_{1-\textit{x}}$Sc$_{\textit{x}}$O$_{3}$
       : Two-component analysis}

\author{Takafumi D. Yamamoto$^{\ast}$}
\author{Ryuji Okazaki}
\author{Hiroki Taniguchi}
\author{Ichiro Terasaki}
\affiliation{Department of Physics, Nagoya University, Nagoya 464-8602, Japan}   %Š'®

\begin{abstract}
We have studied magnetic and transport properties in polycrystalline CaRu$_{1-\textit{x}}$Sc$_{\textit{x}}$O$_{3}$ for 0 $\le \textit{x} \le$ 0.20
in order to clarify the substitution effects of a non-magnetic trivalent ion.
We find that a ferromagnetic transition with $T_{\rm c}$ $\sim $ 30 K is observed in Sc-substituted samples.
The composition dependence of the Curie-Weiss temperature $\theta_{\rm CW}$ implies that the magnetic susceptibility has a paramagnetic contribution
with negative $\theta_{\rm CW}$ and a ferromagnetic contribution with positive $\theta_{\rm CW}$.
The field dependence of magnetization at 2 K is also understood as a summation of the ferromagnetic and paramagnetic components.
These results suggest that CaRu$_{1-\textit{x}}$Sc$_\textit{x}$O$_3$ is a non-uniform magnetic system.
The relationship between the ferromagnetic ordering and the transport properties is also discussed. 
\end{abstract}

\maketitle
\newpage
%%%%%%%%%%%%%%%%%%%%%%%%%%%%%%%%%%%%%%%%%%–{•¶ŠJŽn%%%%%%%%%%%%%%%%%%%%%%%%%%%%%%
%%%%%%%%%%%%%%%%%%%%%%%%%%%%%%%%%%%%%%%%%%Introduction start%%%%%%%%%%%%%%%%%%%%%%%%%%%%%%%%%%%%%%%%
\section{Introduction}
The physical properties of CaRuO$_3$ and SrRuO$_3$ have been extensively studied because of their contrasting magnetism in the same structure.
Both compounds crystallize in an orthorhombic perovskite structure and show a metallic conduction.
On the other hand, they exhibit completely different magnetic properties. 
While SrRuO$_3$ is well known as an itinerant ferromagnet with the transition temperature $T_{\rm c}= 165 $ K,\cite{Shikano,Neumeier}
CaRuO$_3$ shows paramagnetic behavior, in which the magnetic susceptibility obeys the Curie-Weiss law. 
Due to a high Curie-Weiss temperature (-140 K), one can expect an antiferromagnetic ordering in this compound below around 100 K,
but a neutron diffraction and a M\"{o}ssbauer spectroscopy have revealed that
CaRuO$_3$ has no long-range magnetic ordering down to 1.5 K.\cite{Martinez,Gibb} 
As such, the magnetic ground state of CaRuO$_3$ has been controversial, and there are many suggestions in the literature.
Felner \textit{et al.} have found the irreversibility between zero-field-cooled and field-cooled dc magnetization curves and
a magnetic sextet in the M\"{o}ssbauer study for $^{\rm 57}$Fe-substituted CaRuO$_3$ at 4.1 K.\cite{FelnerCaRuO}
From these results, they have proposed that CaRuO$_3$ is not paramagnetic, but rather in a spin-glass state. 
However, the AC susceptibility measurements could not give clear evidence of the spin-glass transition.\cite{Yoshii}
CaRuO$_3$ is believed to be on the verge of a magnetic ordering.
In other words, this compound is readily in a magnetically ordered state resulting from the change of electronic states.
Indeed, 5\%-Sr and 5\%-Na substitutions for Ca induce a spin-glass ordering and
an antiferromagnetic ordering, respectively.\cite{Shepard,Cao}

The substitutions for the Ru site in CaRuO$_3$ have attracted much interest due to their anomalous effects on magnetism. 
It has been reported that the partial substitution of transition metal ions for Ru induces a magnetic ordering.
Surprisingly, this phenomenon occurs regardless of the magnetism of the substituent ions;
a ferromagnetic ordering is reported for $M =$ Fe, Mn, Cr, and Ti
and a spin-glass state for $M =$ Sn, Co, and Cu.\cite{Mizusaki,Kawanaka,Maignan,Hephysrev,CaoSn,Breard,Bradaric}
It should be noted here that ferromagnetism is induced by the substitution of a non-magnetic Ti$^{\rm 4+}$ ion.
This result is incompatible with a simple magnetic dilution effect.
Since this ferromagnetism is induced by only 2 \% of isovalent substitution,
He and Cava have suggested that CaRuO$_3$ is poised at a critical point between ferromagnetism and paramagnetism at low temperature.\cite{Hephysrev}
Recently, Hardy \textit{et al.} have proposed that Ti-substituted CaRuO$_3$ is a heterogeneous itinerant ferromagnetic system,\cite{Hardy}
but the mechanism of the ferromagnetism is still unclear.

A comparison with other ion substitutions would be useful for clarifying the origin of the ferromagnetism. 
We focus on how the magnetism changes with the formal Ru valence.
A trivalent ion substitution creates a pentavalent Ru$^{\rm 5+}$ to increase the formal Ru valence,
which should be compared with the Ti$^{4+}$ substitution.
To see this effect clearly, the substitution of a non-magnetic trivalent ion is desirable, 
because one can exclude a magnetic impurity effects.
However, to our knowledge, there are no such studies up to now.

In this paper, we have measured the magnetization, the electrical resistivity, and the Seebeck coefficient of polycrystalline
CaRu$_{1-\textit{x}}$Sc$_{\textit{x}}$O$_{3}$ (0 $\le \textit{x} \le$ 0.20).
We have found a ferromagnetic order with $T_{\rm c}$ $\sim $ 30 K in Sc-substituted samples.
Based on the assumption that two magnetic components exist in CaRu$_{1-\textit{x}}$Sc$_{\textit{x}}$O$_{3}$,
we discuss a possible picture of the ferromagnetic order.
%%%%%%%%%%%%%%%%%%%%%%%%%%%%%%%%%%%%%%%%%%Introduction end%%%%%%%%%%%%%%%%%%%%%%%%%%%%%%%%%%%%%%%%%%%%%%%%%%%%%%%%%%%%%%%%%%%%%%%%%%%%%%
%%%%%%%%%%%%%%%%%%%%%%%%%%%%%%%%%%%%%%%%%%EXPERIMENTAL DETAILS  start%%%%%%%%%%%%%%%%%%%%%%%%%%%%%%%%%%%%%%%%%%%%%%%%%%%%%%%%%%%%%%%%%%%%%%%%%%%%%%
\section{Experiments}
Polycrystalline samples of CaRu$_{1-\textit{x}}$Sc$_{\textit{x}}$O$_{3}$ (0 $\le \textit{x} \le$ 0.20) were prepared by a conventional solid-state reaction from the stoichiometric mixtures of 
$\rm CaCO_3$\ (99.9\%), $\rm RuO_2$\  (99.9\%), and $\rm Sc_2O_3$\ (99.9\%). 
The mixtures of oxides were ground and heated in air at 1000 $^{\circ}$C for 12 h.
Then, the powders were reground and pressed into pellets, and sintered in air at 1250 $^{\circ}$C for 48 h. 
The prepared samples were investigated by powder x-ray diffraction measurements (Cu K$\alpha$ radiation) at room temperature
 with a Rigaku RINT-2000 diffractometer.

The magnetization measurements were performed by a superconducting quantum interference device magnetometer (Quantum Design MPMS).
Magnetization ($M$) data were collected from 2 to 300 K in the DC field ($H$) of 0.1 T on field cooling and zero field cooling.
The field dependence of magnetization was measured at 2 K in a magnetic field range from -7 to 7 T.
Transport properties were measured from 4.2 to 300 K using homemade probes. 
The electrical resistivity and the Seebeck coefficient were measured using the four-probe technique and the steady-state two-probe method, respectively.
%%%%%%%%%%%%%%%%%%%%%%%%%%%%%%%%%%%%%%%%%%EXPERIMENTAL DETAILS  end%%%%%%%%%%%%%%%%%%%%%%%%%%%%%%%%%%%%%%%%%%%%%%%%%%%%%%%%%%
%%%%%%%%%%%%%%%%%%%%%%%%%%%%%%%%%%%%%%%%%%RESULTS AND DISCUSSION start%%%%%%%%%%%%%%%%%%%%%%%%%%%%%%%%%%%%%%%%%%%%%%%%%%%%%%%%%%
\section{Results and Discussion}
Powder X-ray diffraction patterns of CaRuO$_{3}$ and CaRu$_{0.8}$Sc$_{0.2}$O$_{3}$ are shown representatively in Fig. 1(a),
with a calculated pattern of CaRuO$_{3}$ obtained with RIETAN-FP.\cite{Rietan,Bensch}
The diffraction patterns show that all the prepared samples belong to the orthorhombic perovskite structure of the space group \textit{Pnma}.
Figures 1(b) and 1(c) show the \textit{x} dependence of the lattice constants and the lattice volume $V$ of
CaRu$_{1-\textit{x}}$Sc$_{\textit{x}}$O$_{3}$ at room temperature, respectively. 
The change in the lattice parameter \textit{a} is small, while \textit{b} and \textit{c} increase gradually with Sc content \textit{x}.
Correspondingly, the lattice volume continuously increases with increasing \textit{x}.
We expect that Ru$^{5+}$ ions are generated by Sc substitution and calculate the lattice volume expected from the formula 
Ca[Ru$^{4+}_{1-2 \textit{x}}$Ru$^{5+}_{\textit{x}}$]Sc$^{3+}_{\textit{x}}$O$_{3}$, using the expressions given by 
\begin{eqnarray}
V &=& 224.3 r_{m}^{3} + 173.5 \ \mathrm{[\AA^{3}]},\\
r_{m} &=& (1- 2 \textit{x})r^{4+}_{VI} + \textit{x} r^{5+}_{VI} + \textit{x} r^{3+}_{VI},
\end{eqnarray}
where $r_{m}$ is the average ionic radius and $r^{4+}_{VI}$ (0.62\ \AA), r$^{5+}_{VI}$ (0.565\ \AA), and r$^{3+}_{VI}$ (0.745\ \AA) are the ionic radius of Ru$^{4+}$, Ru$^{5+}$,
and Sc$^{3+}$, respectively.\cite{Taniguchi} 
As shown in Fig. 1(c), the experimentally-observed $V$ is in agreement with the calculated line.
This result suggests successful substitution of Sc for Ru and the existence of Ru$^{5+}$ ions.
A tiny trace (2\%) of CaO was detected for all the samples, which possibly came from evaporation of Ru.
Since CaO is a nonmagnetic insulator, it would cause no significant effects on magnetism. 

Figure 2(a) shows the temperature dependence of $M/H$ in 1 kOe on field cooling for CaRu$_{1-\textit{x}}$Sc$_{\textit{x}}$O$_{3}$.
Typical ferromagnetic behavior is observed below 50 K for the Sc-substituted specimens.
The magnetization below 50 K continues to increase with increasing \textit{x} up to \textit{x} = 0.20. 
We define the Curie temperature $T_{\rm c}$ as the temperature at which the absolute value of $d$($M/H$)/$dT$ takes a maximum
as shown in the inset of Fig. 2(a), and find $T_{\rm c}$ to be 30 K for all the Sc-substituted samples. 
A similar trend has been observed for CaRu$_{1-\textit{x}}$Ti$_\textit{x}$O$_3$.\cite{Hephysrev,Hardy} 
The temperature dependence of $H/M$ is shown in Fig. 2(b). 
A sudden drop corresponding to the onset of the ferromagnetism is found below 50 K for all the samples except for \textit{x} = 0.
For \textit{x} = 0.05, another drop is seen at around 150 K, which was dependent on samples.
Thus the drop near 150 K would be extrinsic.
Above 200 K, $H/M$ linearly increases with temperature.
The linear part of $H/M$ at high temperature is explained using the Curie-Weiss law given by
\begin{equation}
\chi(T) = \frac{C}{T-\theta_{\rm CW}},
\end{equation}
where $C$ is the Curie constant and $\theta_{\rm CW}$ is the Curie-Weiss temperature.
According to this expression, an extrapolation of the linear part to $H/M = 0$ gives a value of $\theta_{\rm CW}$.
The extrapolations are shown as the solid lines for \textit{x} $=$ 0 and 0.20 in Fig. 2(b).
As can be seen in Fig. 2(b), $\theta_{\rm CW}$ shifts to positive with increasing \textit{x}, and the sign finally changes at \textit{x} = 0.20.
We should emphasize that this \textit{x} dependence of $\theta_{\rm CW}$ is seriously incompatible with weakly \textit{x}-dependent $T_{\rm c}$,
because $T_{\rm c}$ is proportional to $\theta_{\rm CW}$ within mean-field approximations in a uniform magnet.
Thus, this result implies that the system is \textit{not} uniform;
the susceptibility should be understood as an average of a paramagnetic part with negative $\theta_{\rm CW}$ and a ferromagnetic part with positive $\theta_{\rm CW}$.
Then the sign change in $\theta_{\rm CW}$ corresponds to the increase in the volume fraction of the ferromagnetic part.
A similar change in $\theta_{\rm CW}$ is seen in Ca$_{1-\textit{x}}$La$_\textit{x}$MnO$_3$,
in which a ferromagnetic interaction is induced by La substitution.\cite{NeumeierApply}
Neumeier and Cohn have pointed out that La substitution causes the magnetic phase segregation in this system
because local ferromagnetic regions appears within the antiferromagnetic phase of CaMnO$_3$.\cite{NeumeierPhysRev}

Based on these speculations, we propose that the magnetic susceptibility for CaRu$_{1-\textit{x}}$Sc$_\textit{x}$O$_3$ has two contributions above $T_{\rm c}$,
given by
\begin{equation}
\chi (\textit{x}, T) = (1- 2 \textit{x}) \chi_p(T) + \textit{x} \chi_{f}(T),
\end{equation}
where $\chi_p (T)$ and $\chi_{f} (T)$ are the paramagnetic and ferromagnetic parts, respectively. 
We assume that the ferromagnetic part consists of the generated Ru$^{5+}$ ions and the paramagnetic one is originated from the Ru$^{4+}$ ions.
We further assume that the volume fraction of the two components is identical to that of the Ru$^{5+}$ and Ru$^{4+}$ ions.
Since the chemical formula is Ca[Ru$^{4+}_{1-2 \textit{x}}$Ru$^{5+}_{\textit{x}}$]Sc$^{3+}_{\textit{x}}$O$_{3}$,
we consider that the paramagnetic phase of CaRuO$_3$ is essentially unaffected with the volume fraction of $1-2 \textit{x}$.
In this assumption, we can regard the paramagnetic part as the magnetic susceptibility of CaRuO$_3$, i.e., $\chi_p (T) = \chi (0,T)$.
Then we can extract the ferromagnetic part $\chi_{f} (T)$ by subtracting the experimentally-observed $\chi(\textit{T})$ of CaRuO$_3$ 
from that of the Sc-substituted samples, using an expression given by 
\begin{equation}
\chi_{f}(T)=\frac{1}{\textit{x}}[\chi(\textit{x},T)-(1- 2 \textit{x})\chi_p(T)].
\end{equation}

Figure 3 shows the temperature dependence of thus obtained $\chi_{f}$.
All the data fall into a single curve,
supporting the validity the two-component model described by Eq. (4).
A broken curve depicts the calculation given by
\begin{equation}
\chi_{f}(T) = \frac{C_{f}}{T-T_{\rm c}},
\end{equation}
where $C_{f}$ is the Curie constant of the ferromagnetic component and $T_{\rm c}$ is the ferromagnetic transition temperature.
The calculated curve with $C_{f} =$ 2.08 emu K/mol and $T_{\rm c} =$ 30 K explains well the experimental data.
We employed the value of $T_{\rm c}$ from Fig. 2(a).
From the value of $C_{f}$, we obtain an effective magnetic moment $\mu_{\rm eff}$ to be 4.08$\mu_{\rm B}$ per formula unit, corresponding to $S \sim$ 3/2.
This result implies that Ru$^5{+}$ ions with $S =$ 3/2 constitute the ferromagnetic component.

To further investigate the two-component analysis, the field dependence of magnetization was measured.
We have examined that a demagnetization correction is negligible for all the measurements.
Figure 4(a) shows the $M$-$H$ curve at 2 K for CaRu$_{1-\textit{x}}$Sc$_{\textit{x}}$O$_{3}$.
CaRuO$_3$ exhibits a linear field dependence, showing the absence of magnetic ordering.
On the other hand, magnetic hysteresis loops are observed at 2 K in the Sc-substituted samples, which is indicative of ferromagnetism.
With increasing Sc, the ferromagnetic hysteresis loop becomes more remarkable and the magnetization increases, reflecting the development of the ferromagnetic order.
Note that $M$ increases linearly above 60 kOe without any sign of saturation.
This feature is also observed in previous studies of other transition metal ion substitution for Ru.\cite{Maignan,HeCondens,Hardy}
We assume that $M$-$H$ curve can be understood as a summation of the ferromagnetic and paramagnetic contribution, as was already discussed above.
Then the experimentally-observed magnetization $M(\textit{x}, H)$ is given by
\begin{equation}
M(\textit{x}, H) =(1- 2 \textit{x})M_p (H) + \textit{x}M_{f} (H),
\end{equation}
where $M_{f}(H)$ and $M_p (H) = \chi_p H$ are the ferromagnetic and paramagnetic parts, respectively. 
Accordingly, the ferromagnetic component $M_{f} (H)$ is given by
\begin{equation}
M_{f} (H) = \frac{1}{\textit{x}}[M(\textit{x}, H)-(1- 2\textit{x})M_p (H)].
\end{equation}

The field dependence of $M_{f}$ is shown in Fig. 4(b).
All the data almost fall into a single curve again, and this scaling further confirms the validity of the two-component model.
We evaluate the saturation magnetization $M_{s}$ to be about 1.0$\mu_{\rm B}$ per formula unit by extrapolating $M_{f}$ above 60 kOe to $H =$ 0. 
This value is only one-third of $S=$ 3/2.

Figure 5(a) shows the temperature dependence of the electrical resistivity $\rho$.
The resistivity shows metallic conduction ($d \rho/dT>$ 0) for CaRuO$_3$, while it systematically changes to  non-metallic one ($d \rho/dT<$ 0)
with increasing Sc content.
For example, $\rho$ continues to increase by two orders of magnitude from 300 down to 4.2 K for CaRu$_{0.80}$Sc$_{0.20}$O$_3$. 
We should note that the resistivity increases sharply below 50 K, as seen in the change in the slope of the solid lines in Fig. 5(a).
One can attribute the rapid increase in $\rho$ to magnetic domain boundaries.\cite{Ju,Kobayashi}
When spin-polarized conduction electrons in one domain move to the neighboring domains, 
they undergo the spin-dependent scattering at the boundaries between domains with different magnetization directions.
The temperature dependence of the Seebeck coefficient $S$ is shown in Fig. 5(b).
In contrast to the resistivity, the Seebeck coefficient retains metallic against Sc substitution. 
Since the diffusive term of the Seebeck coefficient is predominantly determined by carrier concentration,
the Sc substitution mainly affects not the carrier concentration but the mobility. 
Moreover, one can see that $S$ increases with \textit{x} more remarkably at low temperatures, while it is almost independent near room temperature.

To see this clearly, we plot $\Delta S = S(\textit{x}) - S (\textit{x} = 0)$ in Fig. 6, where 
$\Delta S$ takes a maximum near the transition temperature $T_{\rm c}$ and the maximum increases with increasing Sc content \textit{x}. 
The shape of $\Delta S$ reminds us of the magnetic susceptibility near $T_{\rm c}$,
suggesting that the spin fluctuation affects the transport properties.
It has been theoretically suggested that the spin fluctuation influences the transport coefficients in high-$T_{\rm c}$ copper oxides.\cite{Kontani}
The relationship between the ferromagnetic order and the transport properties is indicated in a qualitative manner,
and a further study such as magneto-transport needs to be performed.

Finally, we discuss the nature of the ferromagnetism.
We have shown that CaRu$_{1-\textit{x}}$Sc$_\textit{x}$O$_3$ is a non-uniform magnetic system,
consisting of the paramagnetic component originated from CaRuO$_3$ and the ferromagnetic one driven by Sc substitution.
Considering the identical $T_{\rm c} =$ 30 K, a phase segregation is likely to occur.
He and Cava have pointed out that the substitution of 3\textit{d} transition metals \textit{M} drives CaRu$_{1-\textit{x}}$M$_\textit{x}$O$_3$
to an inhomogeneous ferromagnetic system, containing ferromagnetic clusters with an intrinsic $T_{\rm c}$.\cite{HeCondens}
The Sc substitution effect could be understood by their picture.
However, the domain size, the distribution pattern, and the segregation mechanism are to be explored.
The value of $\mu_{\rm eff}$ above $T_{\rm c}$ indicates that Ru$^{5+}$ ions generate the ferromagnetic component.
One open issue is that the saturation magnetization is only one-third of $S =$ 3/2.
One possibility is that only one-third of Ru$^{5+}$ ions contribute to the ferromagnetic order.
In this case, the rest of Ru$^{5+}$ ions would show paramagnetic behavior and result in the linearly increase of $M_f$ at high field.
A second possibility is that the ferromagnetism is of itinerant nature.
The discrepancy between $\mu_{\rm eff}$ and $M_{\rm s}$ is often observed in itinerant ferromagnets,\cite{Grewe}
which is consistent with the moderately conductive behavior.
At least we can say that the ferromagnetism observed in the present work is different from that in SrRuO$_3$
in the sense that the generated Ru$^{5+}$ ions are responsible for the order.
At the present stage, the magnetism of CaRu$_{1-\textit{x}}$Sc$_\textit{x}$O$_3$ is still enigmatic in some respects.
To clarify the picture of the magnetism in more detail,
the microscopic investigations such as transmission microscope in the real space and neutron scattering in the momentum space are indispensable.
%%%%%%%%%%%%%%%%%%%%%%%%%%%%%%%%%%%%%%%%%%RESULTS AND DISCUSSION end%%%%%%%%%%%%%%%%%%%%%%%%%%%%%%%%%%%%%%%%%%%%%%%%%%%%%%%%%%
%%%%%%%%%%%%%%%%%%%%%%%%%%%%%%%%%%%%%%%%%%Summary start%%%%%%%%%%%%%%%%%%%%%%%%%%%%%%%%%%%%%%%%%%%%%%%%%%%%%%%%%%
\section{Summary}
We have measured magnetization and the transport coefficients of polycrystalline CaRu$_{1-\textit{x}}$Sc$_\textit{x}$O$_3$ (0 $\le$ \textit{x} $\le$ 0.20).
The ferromagnetic order with $T_{\rm c} \sim$ 30 K is found in all the Sc-substituted samples.
The Curie-Weiss temperature $\theta_{\rm CW}$ shows the anomalous composition dependence,
implying that the susceptibility has a paramagnetic contribution with negative $\theta_{\rm CW}$ and a ferromagnetic contribution with positive $\theta_{\rm CW}$.
We actually find that the $M/H$ - $T$ and $M$-$H$ curves can be understood as a summation of the ferromagnetic and paramagnetic components.
Based on these results, we conclude that CaRu$_{1-\textit{x}}$Sc$_\textit{x}$O$_3$ is a non-uniform magnetic system
consisting of the paramagnetic component originated from CaRuO$_3$ and the ferromagnetic one driven by Sc substitution.
Furthermore, we have found that the transport properties are substantially affected by the ferromagnetic order.

%%%%%%%%%%%%%%%%%%%%%%%%%%%%%%%%%%%%%%%%%%Summary end%%%%%%%%%%%%%%%%%%%%%%%%%%%%%%%%%%%%%%%%%%%%%%%%%%%%%%%%%%

%%%%%%%%%%%%%%%%%%%%%%%%%%%%%%%%%%%%%%%%%%Acknowledgement start%%%%%%%%%%%%%%%%%%%%%%%%%%%%%%%%%%%%%%%%%%%%%%%%%%%%%%%%%%
\section*{Acknowledgements}
\addcontentsline{toc}{section}{ACKNOWLEDGMENTS}
This work is partially supported by a Grant-in-Aid for Scientific Research, MEXT, Japan (Nos. 25610091, 26247060).

%%%%%%%%%%%%%%%%%%%%%%%%%%%%%%%%%%%%%%%%%%%figure%%%%%%%%%%%%%%%%%%%%%%%%%%%%%%%%%%%%%%%%%%%%%%%%%%%%%%%%%%%%%%%
%\newpage
%%%%%%%%%%%%%%%%%%%%%%%%%%%%%%%%%%%%%%%%%%Acknowledgement end%%%%%%%%%%%%%%%%%%%%%%%%%%%%%%%%%%%%%%%%%%%%%%%%%%%%%%%%%%

\newpage
\begin{figure}[htp]
 \begin{center}
  \includegraphics[scale=0.4,clip]{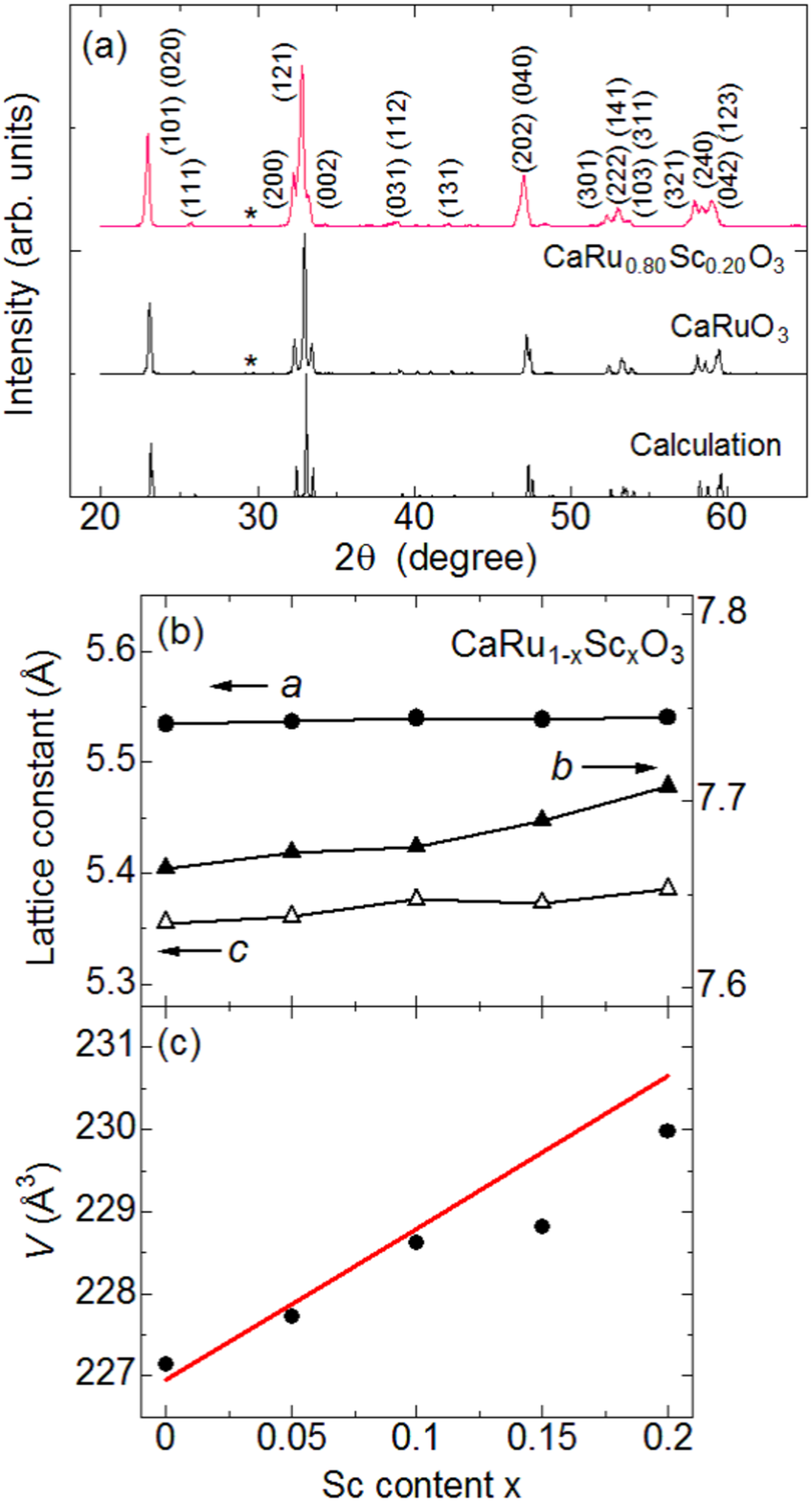}
   \caption{(Color online) (a) The powder X-ray diffraction patterns of CaRuO$_{3}$ and CaRu$_{0.8}$Sc$_{0.2}$O$_{3}$ at room temperature.
	The calculated pattern of CaRuO$_{3}$ is shown at the bottom. The asterisk shows an reflection of CaO.
	(b) The lattice constants for \textit{a}, \textit{c} (left scale), and \textit{b} axis (right scale) and (c) the lattice volume $V$ 
 of CaRu$_{1-\textit{x}}$Sc$_{\textit{x}}$O$_{3}$ as a function of Sc content \textit{x}, respectively. 
 A solid line in (c) depicts the calculation expected from the formula 
 Ca[Ru$^{4+}_{1-2 \textit{x}}$Ru$^{5+}_{\textit{x}}$]Sc$^{3+}_{\textit{x}}$O$_{3}$ (see text).}
 \end{center}
\end{figure}

\begin{figure}[htp]
 \begin{center}
  \includegraphics[scale=0.4,clip]{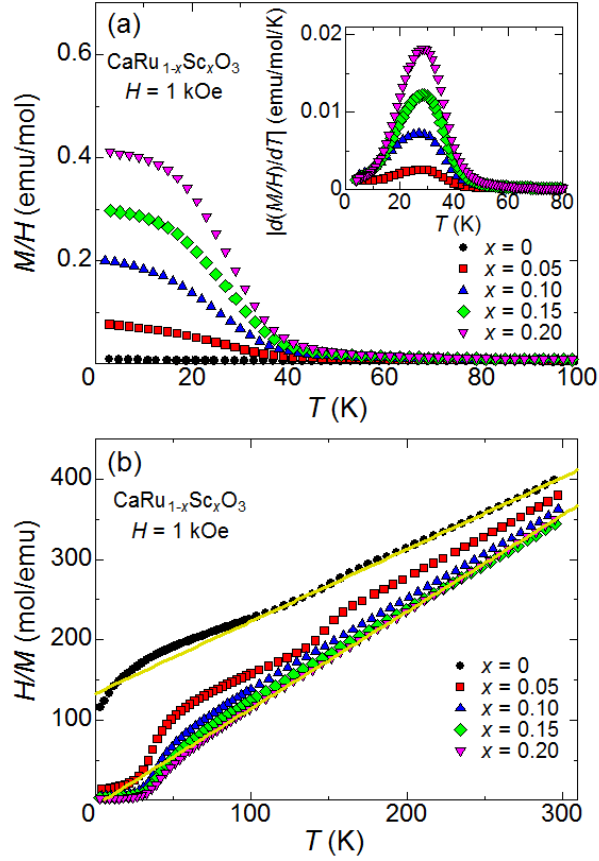}
   \caption{(Color online) Temperature dependence of (a) $M/H$  and (b) $H/M$ measured at 1 kOe on field cooling
 for CaRu$_{1-\textit{x}}$Sc$_{\textit{x}}$O$_{3}$, respectively.
 The inset in (a) shows temperature dependence of $d$($M/H$)/$dT$.
 The solid lines in (b) depict an extrapolation of the linear part to $H/M = 0$ (see text).}
 \end{center}
\end{figure}

\begin{figure}[htp]
 \begin{center}
  \includegraphics[scale=0.4,clip]{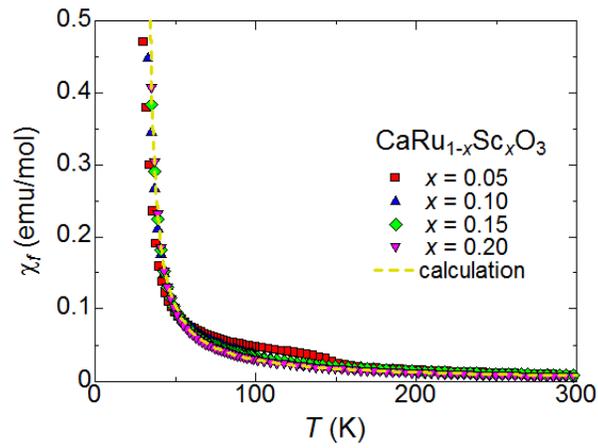}
   \caption{(Color online) Temperature dependence of a ferromagnetic component $\chi_{f}$ for CaRu$_{1-\textit{x}}$Sc$_{\textit{x}}$O$_{3}$.
A broken line depicts the calculation derived from $\chi_{f} = C_{f}/ (T-T_{\rm c})$ (see text).}
 \end{center}
\end{figure}

\begin{figure}[htp]
 \begin{center}
  \includegraphics[scale=0.4,clip]{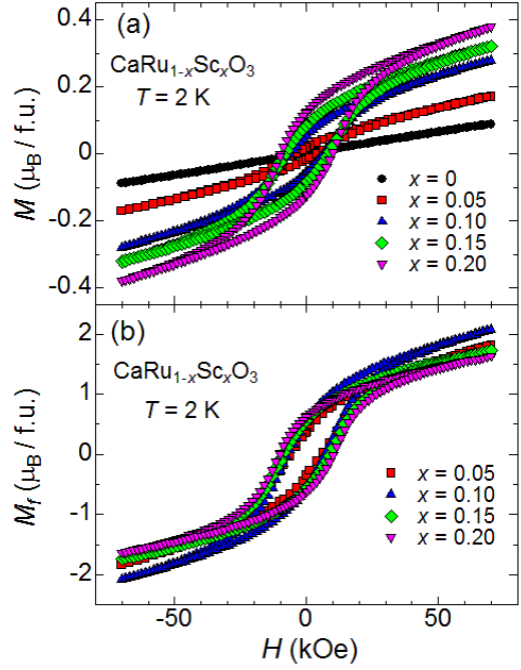}
   \caption{(Color online) Magnetic-field dependence of (a) $M$ and (b) $M_{f}$ (see text) measured
            at 2 K for CaRu$_{1-\textit{x}}$Sc$_\textit{x}$O$_3$.}
 \end{center}
\end{figure}

\begin{figure}[htp]
 \begin{center}
  \includegraphics[scale=0.4,clip]{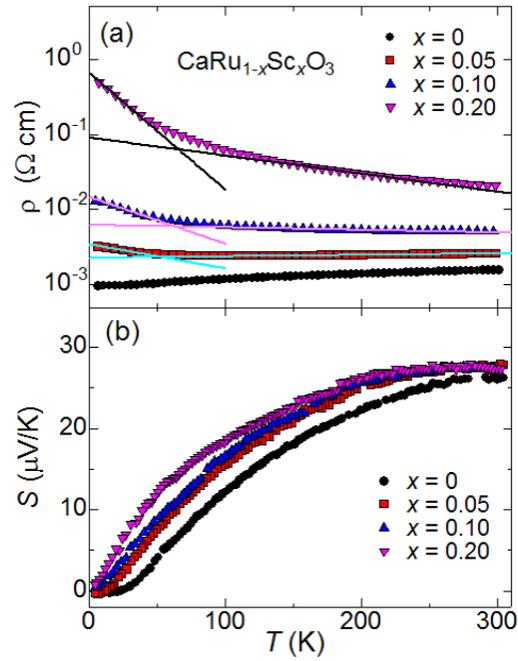}
   \caption{(Color online) Temperature dependence of (a) the electrical resistivity and (b) the Seebeck coefficient
for CaRu$_{1-\textit{x}}$Sc$_\textit{x}$O$_3$, respectively.
The solid lines are guides to the eye to emphasize the slope change.}
 \end{center}
\end{figure}

\begin{figure}[htp]
 \begin{center}
  \includegraphics[scale=0.4,clip]{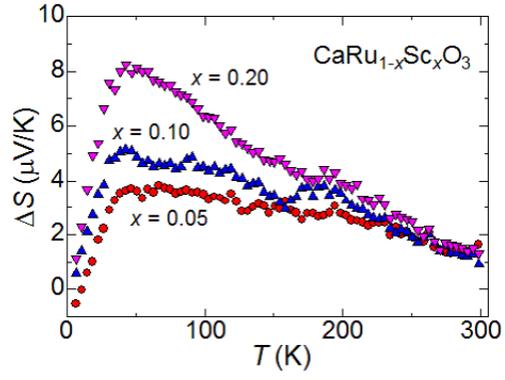}
   \caption{(Color online) Temperature dependence of $\Delta S = S (x)- S (x = 0)$ for CaRu$_{1-\textit{x}}$Sc$_\textit{x}$O$_3$.}
 \end{center}
\end{figure}

\end{document}